# Stabilization of spatiotemporal solitons in Kerr media by dispersive coupling

Yaroslav V. Kartashov,[1,2,*] Boris A. Malomed,[3] Vladimir V. Konotop,[4] Valery E. Lobanov,[5] and Lluis Torner[1]

[1]*ICFO-Institut de Ciencies Fotoniques, and Universitat Politecnica de Catalunya, Mediterranean Technology Park, 08860 Castelldefels (Barcelona), Spain*
[2]*Institute of Spectroscopy, Russian Academy of Sciences, Troitsk, Moscow Region, 142190, Russia*
[3]*Department of Physical Electronics, School of Electrical Engineering, Faculty of Engineering, Tel Aviv University, Tel Aviv 69978, Israel*
[4]*Centro de Física Teórica e Computacional and Departamento de Física, Faculdade de Ciências, Universidade de Lisboa, Avenida Professor Gama Pinto 2, Lisboa 1649-003, Portugal*
[5]*Russian Quantum Center, Skolkovo, Moscow Region, 143025, Russia*

We introduce a mechanism to stabilize spatiotemporal solitons in Kerr nonlinear media, based on the dispersion of linear coupling between the field components forming the soliton states. Specifically, we consider solitons in a two-core guiding structure with inter-core coupling dispersion (CD). We show that CD profoundly affects properties of the solitons, causing the *complete stabilization* of the otherwise highly unstable spatiotemporal solitons in Kerr media with focusing nonlinearity. We also find that the presence of CD stimulates the formation of bound states, which however are unstable.

The stability of solitons is a problem of fundamental significance in optics and studies of atomic Bose-Einstein condensates (BECs) [1-3]. A general problem is that in a uniform medium with cubic nonlinearity, induced by the Kerr effect in optics and inter-atomic collisions in BEC, two- and three-dimensional (2D and 3D) soliton solutions are highly unstable due to the occurrence of the critical and supercritical collapse in the 2D and 3D settings, respectively [4,5]. In particular, the 2D nonlinear Schrödinger equation (NLSE) equation gives rise to the so-called Townes soliton (TS) [6], which is *degenerate* in free space, in the sense that it occurs only for a single value of the energy flow. On physical grounds, the TS represents an unstable state that separates the regimes of light spreading caused by diffraction and non-arrested self-focusing, termed beam collapse [4,5]. Over the years, physical systems governed by the two-dimensional NLSEs coupled by nonlinear or linear terms have been studied, giving rise to two- or multi-component versions of the TS, which are unstable too [7].

A variety of approaches to cope with the problem and thus stabilize multidimensional soliton propagation in materials with Kerr nonlinearity have been suggested. Among them is the utilization of external linear potentials [8,9], inhomogeneous nonlinearities [10,11], addition of competing nonlinear terms [12] and nonlocality of the material response [13,14], longitudinal modulations of the material parameters [15], as well as inclusion of various dissipative effects, to name just a few approaches. Nevertheless, elucidation of physical mechanisms that can be implemented in practice remains an important task, as, because of the universality of the Kerr nonlinear response, it may open the way for creation of stable excitations in a large variety of physical settings. Recently, an indication of the existence of a new stabilization mechanism was reported in Ref [16], for a binary BEC with linear spin-orbit coupling between two components. Such coupling enters the governing equations through first-order spatial derivatives of the wave functions [17,18]. Two-dimensional soliton complexes in such a system of coupled NLS equations were shown to be stable as the derivative linear coupling lifts the degeneracy, thus protecting them from the collapse-driven instability. Generally, the spin-orbit coupling has been shown to allow formation of a variety of matter-wave solitons [19-23].

In this Letter we show that a related general mechanism, but stemming from a different physical effect, viz., temporal dispersion of the linear coupling between field components, stabilizes spatiotemporal optical solitons in Kerr media as well. As an example of the scheme where the mechanism can be realized, we consider the soliton propagation in a focusing Kerr material with anomalous group-velocity dispersion (GVD), structured as a dual-core planar waveguide, with shallow modulation of the refractive index only along one of the transverse axes $(y)$. In the framework of coupled-mode theory, the propagation of light in such a planar dual-core waveguide is governed by two coupled NLSEs for the scaled field amplitudes $q_{1,2}$ in the two cores of the guiding structure:

$$i\frac{\partial q_1}{\partial \xi} = -\frac{1}{2}\left(\frac{\partial^2}{\partial \eta^2} + \frac{\partial^2}{\partial \tau^2}\right)q_1 - |q_1|^2 q_1 - \left(c + i\delta\frac{\partial}{\partial \tau}\right)q_2 - \beta q_1,$$
$$i\frac{\partial q_2}{\partial \xi} = -\frac{1}{2}\left(\frac{\partial^2}{\partial \eta^2} + \frac{\partial^2}{\partial \tau^2}\right)q_2 - |q_2|^2 q_2 - \left(c + i\delta\frac{\partial}{\partial \tau}\right)q_1 + \beta q_2. \quad (1)$$

Here the transverse coordinate $\eta = x/x_0$ and the propagation distance $\xi = z/L_{\text{dif}}$ are normalized, respectively, to a characteristic spatial scale $x_0$ and diffraction length

$L_{\text{dif}} = k x_0^2$, where $k$ is the wavenumber, and $\tau = (t - v_{\text{gr}}^{-1} z)/t_0$ is the normalized reduced time, with characteristic temporal scale $t_0 = (L_{\text{dif}} |\partial^2 k/\partial \omega^2|)^{1/2}$ determined by the GVD, $\partial^2 k/\partial \omega^2$. The strength of the linear coupling between the cores, $c = (k L_{\text{dif}}/n) \int \delta n_1 w_1 w_2 d\zeta / \int w_1^2 d\zeta$, is determined by the overlap integral between the modes $w_{1,2}(\zeta)$ of the individual planar waveguides and by the refractive index variation $\delta n_{1,2}(\zeta) \ll n$ inside the individual waveguides, in the direction of $\zeta = y/x_0$ perpendicular to the cores; the coefficient $\beta$ in Eq. (1) accounts for a possible refractive-index mismatch between the cores; finally, $\delta \approx (c/t_0)[2/\omega + \chi_1^{-1} \partial \chi_1/\partial \omega]$, where $\chi_1$ is the amplitude of the permittivity modulation, characterizes the first-order coupling dispersion (CD), introduced for dual-core fibers in [24,25]. Naturally, the CD becomes more pronounced as the pulse width decreases. From the experimental point of view, realization of sufficiently strong CD necessary for soliton stabilization may be achieved by proper engineering of the geometry of coupled waveguides. The field amplitudes in Eq. (1) are defined as $|q_{1,2}|^2 = (k n_2 L_{\text{dif}}/n) |E_{1,2}|^2 \int w_{1,2}^4 d\zeta / \int w_{1,2}^2 d\zeta$, where $n_2$ is the nonlinear refractive index, with the nonlinear contribution to the refractive index in each core determined solely by the field amplitude in the same core. One can see that Eq. (1) describes the propagation of effectively two-dimensional excitations that are allowed to diffract along the spatial axis $\eta$ and disperse in time $\tau$, while the field structure along direction $\zeta$, perpendicular to the cores, is represented by amplitudes $q_{1,2}$. The parameters of the medium do not vary along $\eta$ or $\tau$.

In addition to the energy flow, $U = \iint (|q_1|^2 + |q_2|^2) d\eta d\tau$, and the momentum, Eqs. (1) conserve the Hamiltonian,

$$H = \iint [\beta(|q_2|^2 - |q_1|^2) - c(q_1 q_2^* + q_2 q_1^*) - i\delta(q_1^* \partial q_2/\partial \tau + q_2^* \partial q_1/\partial \tau) + \sum_{n=1,2} (1/2)(|\partial q_n/\partial \eta|^2 + |\partial q_n/\partial \tau|^2 - |q_n|^4)] d\eta d\tau. \quad (2)$$

We search for soliton solutions of Eq. (1) with the form $q_{1,2}(\tau, \eta, \xi) = w_{1,2}(\tau, \eta) \exp(i b \xi)$, with complex stationary functions $w_{1,2} \equiv \text{Re}(w_{1,2}) + i \text{Im}(w_{1,2})$. To understand the conditions required for the formation of stationary solutions, we first consider the spectrum of the linear version of Eq. (1). Letting $q_{1,2} = a_{1,2} \exp(i k_\eta \eta - i \omega \tau + i b \xi)$, where $k_\eta$ and $\omega$ are the wavenumber and frequency, and $b$ is the propagation constant, one arrives at the linear dispersion relation:

$$b_\pm = -(1/2)(k_\eta^2 + \omega^2) \pm [\beta^2 + (c + \delta \omega)^2]^{1/2}. \quad (3)$$

The dependence $b_\pm(k_\eta, \omega)$ is symmetric with respect to the $k_\eta$, but asymmetric with respect to $\omega$. The localized nature of the soliton solutions requires that their propagation constant belongs to the region $b \geq b_{\text{co}}(\beta, c, \delta) = \max[b_+(k_\eta, \omega)]$, where we take into account that $b_+ > b_-$ in Eq. (3). Thus, a localized soliton solution should bifurcate from the upper edge $b_+$ of the spectrum (3) at $k_\eta = 0$ and $\omega \neq 0$, if $c \neq 0$. This suggests that the phase front of the soliton is tilted along the $\tau$-axis. The upper edge of the linear spectrum monotonically grows with the increase of mismatch $\beta$ or of CD $\delta$. We set $c = 1$, unless stated otherwise.

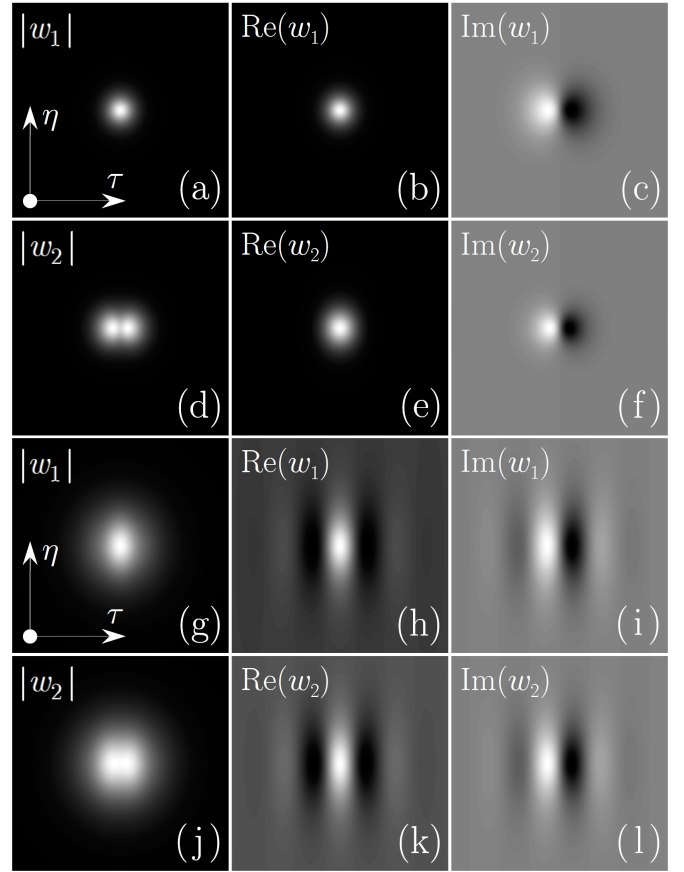

Fig. 1. Absolute value and real and imaginary parts of the field of fundamental solitons for $\delta = 1$ (a)-(f) and $\delta = 2.5$ (g)-(l) at $b = 5$, $\beta = 3$. The panels show the $\tau, \eta \in [-5, +5]$ window and the solutions correspond to the circles in Fig. 2(b).

We are interested in soliton solutions carrying most of the energy flow in the first core of the guiding structure (i.e., in the component $q_1$), which are similar to so-called slow vector states, which typically feature enhanced stability in comparison with other soliton branches in bimodal systems [26-29]. Representative examples of fundamental soliton shapes are shown in Fig. 1. In the general case, with $c, \delta \neq 0$, both components $w_{1,2}$ exhibit a nontrivial phase structure. The real parts of the field are even in $\tau$ and $\eta$, while the imaginary parts are even in $\eta$, but odd in $\tau$. The field phases are close to the linear functions $\phi_{1,2} \equiv \arg(w_{1,2}) \approx -\Omega_{1,2} \tau$. Noticeable deviations from this linear phase profile are observed only for relatively small CD, $\delta < 1$, around intensity maxima of soliton solutions. The frequencies $\Omega_{1,2}$ greatly increase with $\delta$, causing pronounced oscillations in the real and imaginary parts of the field [compare Figs. 1(c) and 1(i)]. An increase of $\delta$ at a fixed $b$, leads to a progressive broadening of the soliton solutions, and a decrease of their amplitude. While the profile of the absolute value of the field in the first core is always bell-shaped [as in Fig. 1(a)], in the second core one can see two bright spots, which represent a specific structure of the $q_2$ field imposed by the CD [Fig. 1(d)]. Note that at $c = 0$, in the absence of the constant part of the inter-core coupling, one may construct soliton solutions in which the only non-vanishing components are $\mathrm{Re}(w_1)$ and $\mathrm{Im}(w_2)$.

Typical dependences of the energy flow $U$ of the stationary solutions versus their propagation constant and the value of CD are presented in Fig. 2, for both $c=1$ and $c=0$. The dependence $U(b)$ for $c=1$ is non-monotonous [Fig. 2(a)]. One observes that $U$ increases at $b \to \infty$, but it rapidly grows when $b$ approaches cutoff $b_{\mathrm{co}}$ (for $b$ close to $b_{\mathrm{co}}$, $U$ substantially exceeds the TS energy flow, i.e. $U_\mathrm{T} \approx 5.85$). The soliton amplitude vanishes at cutoff, while its width monotonically grows at $b \to b_{\mathrm{co}}$ [in particular, $b_{\mathrm{co}} = \beta$ at $c=0$, see Eq. (3)]. The larger the value of the propagation constant, the more pronounced the soliton localization.

The plots show that soliton solutions exist above a threshold value of $U$ that may be substantially lower than $U_\mathrm{T}$. Thus, inclusion of the CD reduces the soliton existence threshold. In Fig. 2(a) we also observe that, as $b$ increases, $U$ asymptotically approaches $U_\mathrm{T}$. This is explained by the strong imbalance of the energy flows in the two cores, $U_1 \gg U_2$, due to a large mismatch of the respective propagation constants, $b+\beta$ and $b-\beta$. The imbalance is an important property, which makes it possible to obtain analytical results for the stationary solitons, as shown below. The stability/instability transition in Fig. 2(a) only partly complies with the Vakhitov-Kolokolov criterion $dU/db \geq 0$ [4,5].

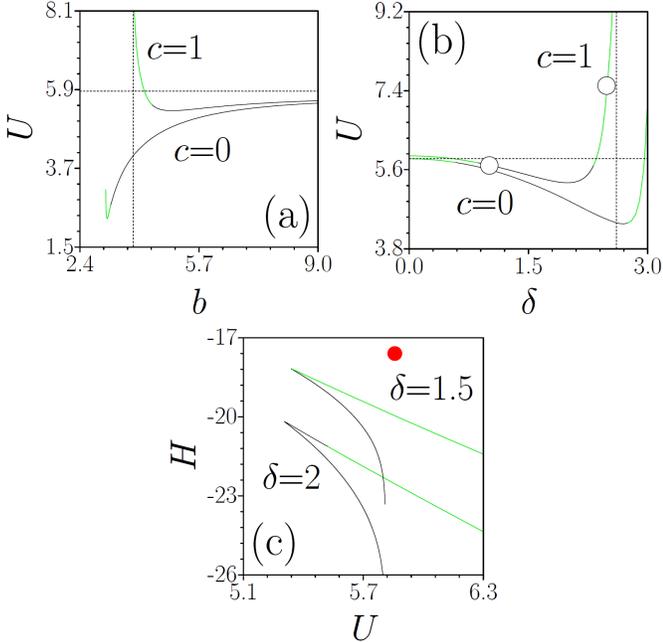

Fig. 2. (a) Energy flow of fundamental solitons versus (a) the propagation constant for $\beta=3$, $\delta=2$, and (b) the CD for $b=5$, $\beta=3$. Horizontal dashed lines indicate the energy flow of the TS $U_\mathrm{T} \approx 5.85$ [4,5], while vertical dashed lines indicate the boundary of the linear spectrum at $c=1$. Circles in (b) correspond to solitons in Fig. 1. (c) Hamiltonian versus energy flow for $\beta=3$, $c=1$. The red dot stands for the TS, at $\beta=3$, $c=0$, $\delta=0$. Stable and unstable soliton branches are shown by black and green lines, respectively. Families for all values of $\delta$ asymptotically approach the energy flow corresponding to the TS.

Figure 2(b) shows the dependence $U(\delta)$ for a fixed propagation constant. This dependence is non-monotonous too. An increase of the CD initially causes a decrease of $U$, starting from $U_\mathrm{T}$ at $c=\delta=0$ (which is explained below analytically), but at $\delta \to \delta_{\mathrm{co}}$ one is again approaching the edge of the linear spectrum, that leads to broadening of the soliton and rapid growth of $U$. Figure 2(c) shows that, unlike the TS, which is represented by an isolated dot on the Hamiltonian–energy flow plane, in the case of nonzero CD one obtains continuous $H(U)$ families. Such dependences can be used for drawing conclusions about soliton stability [30]. The upper branches of $H(U)$ dependencies corresponding to the regions where $dU/db<0$, are nearly linear. While such upper branches do not realize the lowest value of the Hamiltonian for a given value of the energy flow, thus may be viewed as excited states, a small portion of them was found to correspond to stable propagation too.

As mentioned above, the two-component soliton solutions tend to be strongly asymmetric, i.e. $|q_2| \ll |q_1|$. Approximate analytical solutions can be obtained for such strongly asymmetric states, assuming that $b$ and $\beta$ are large enough, while the difference $\Delta b = b-\beta > 0$ is relatively small. In the zero-order approximation, the solution to the stationary version of Eq. (1) is $w_1 = w_\mathrm{T}(\rho)$, $w_2 = 0$, where $\rho^2 = \eta^2 + \tau^2$ and $w_\mathrm{T}(\rho)$ is the standard TS [4,5]. In the next approximation, one gets $w_2 = (2\beta)^{-1}(cw_\mathrm{T} + i\delta\tau\rho^{-1}dw_\mathrm{T}/d\rho)$ and $\mathrm{Im}(w_1) = -c\delta\beta^{-1}\tau w_\mathrm{T}$. At the same order, the equation for a correction to $\mathrm{Re}(w_1)$ is

$$\left[2\Delta b - \left(\frac{\partial^2}{\partial \eta^2} + \frac{\partial^2}{\partial \tau^2}\right) - 6w_\mathrm{T}^2\right]\mathrm{Re}(w_1^{(1)}) = \frac{c^2}{\beta}w_\mathrm{T} - \frac{\delta^2}{\beta}\frac{\partial^2 w_\mathrm{T}}{\partial \tau^2} \quad (4)$$

An exact solution $\mathrm{Re}(w_1^{(1)}) = (2\beta\rho)^{-1}(\delta\tau)^2 dw_\mathrm{T}/d\rho$ to Eq. (4) can be obtained when $c=0$, which yields the corresponding correction to the soliton energy flow, $U^{(1)} = 2\iint w_\mathrm{T} \mathrm{Re}(w_1^{(1)}) d\eta d\tau = -(2\beta)^{-1}\delta^2 U_\mathrm{T}$. Comparing it with the numerical result displayed in Fig. 2(b) for $c=0$, we see that it explains the parabolic dependence $U(\delta)$ at small $\delta$, the respective coefficient being predicted with a relative error $\simeq 20\%$. Increasing $b$ and $\beta$ by a factor of 2 reduces the error to $\simeq 10\%$.

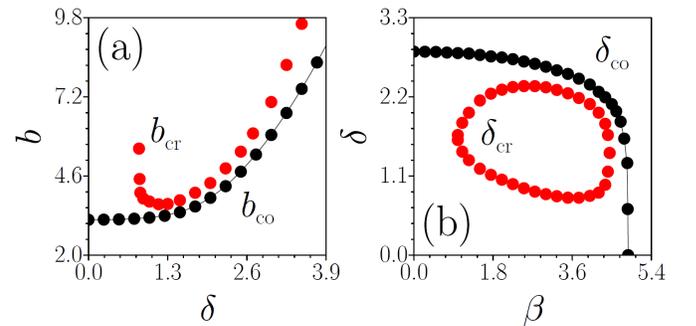

Fig. 3. Existence and stability domains for the soliton solutions in the $(\delta, b)$ plane for $\beta=3$ (a), and in the $(\beta, \delta)$ plane for $b=5$ (b). In (a), solitons exist when $b > b_{\mathrm{co}}$, and are stable when $b > b_{\mathrm{cr}}$. In (b), they exist when $\delta < \delta_{\mathrm{co}}$, being stable in the domain bounded by the red dots and labeled $\delta_{\mathrm{cr}}$. In both plots, $c=1$.

Therefore, as readily visible in Fig 2(c), the central result of this Letter is that the CD leads to the formation of wide families of stable two-dimensional spatiotemporal solitons. The stability of the soliton solutions was tested by adding white-noise perturbations (typically $0.1\%$ in the amplitude) to the initial conditions and calculating numerically their subsequent evolution up to $\xi = 500$. Solitons that kept their shape in the course of the propagation were classified as stable ones. At both $c=1$ and $c=0$, only the low-amplitude branch close to the propagation-constant cutoff was found to be unstable in Fig. 2(a), with the entire family becoming stable above a critical value of the propagation constant, $b > b_{cr}$. The dependencies $U(\delta)$ reveal two instability domains in Fig. 2(b). As expected, the soliton solutions are unstable when CD $\delta$ is small. Increasing $\delta$ results in stabilization of the soliton solution, but stability is again lost in a part of the branch with $\partial U / \partial \delta < 0$, as one approaches the edge of the linear spectrum and as a consequence the soliton amplitude decreases. A summary of the stability domains is displayed in Fig. 3 in the parameter planes $(\delta,b)$, for a fixed mismatch $\beta$, and $(\beta,\delta)$, for a fixed propagation constant $b$. In Fig. 3(a) the stability boundary sharply rises close to $\delta \approx 0.83$, and no stability was found when CD was too small. In Fig. 3(b), soliton solutions exist below the $\delta_{co}$ curve, which is defined by the boundary of the linear spectrum [see Eq. (3)]. In that case, stable solutions were found only in a limited range of the CD coefficient and in a limited range of values of the refractive-index mismatch. Thus, one concludes that an inter-core mismatch is necessary for stabilization to occur, and thus no stable solutions exist when $\beta = 0$. Note that at $\beta \to b$, the first component $q_1$ strongly broadens and solutions become unstable, in agreement with the general instability trend for low-amplitude broad excitations, for which the CD stabilization effect is negligible.

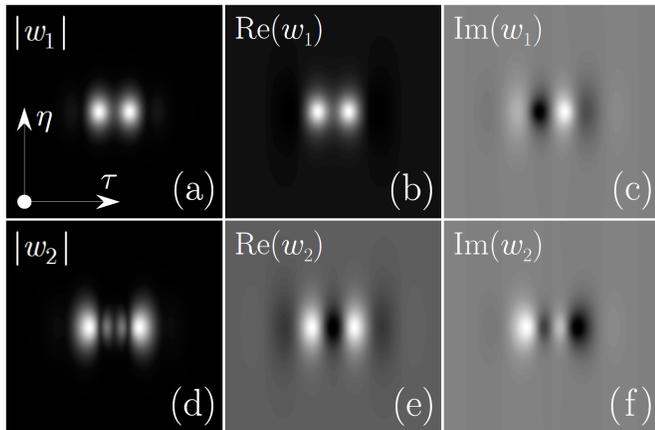

Fig. 4. The absolute value and real and imaginary parts of the field for dipole solitons with $b=9$, $\beta=3$, $\delta=3.4$, $c=1$. The plots are shown within $\eta,\tau \in [-3,+3]$ window.

Finally, we found that the presence of a suitable CD not only affects the stability of fundamental soliton solutions, but that it also makes formation of their stationary bound states possible. An example of a dipole-mode bound-state is shown in Fig. 4. The bright spots in such bound states were always found to be aligned along the $\tau$ axis, and their phase structure was found to be rich. In particular, in some cases, the phase of the component $q_2$ was observed to exhibit rapid variations around $\tau = 0$ on top of an almost constant phase gradient typical also for fundamental solitons. We found that bound states require a minimum value of the CD coefficient $\delta$ for their existence. As expected for such excited states all dipole solutions were found to be unstable upon propagation.

Summarizing, we have shown that the CD (coupling dispersion) dramatically modifies properties of spatiotemporal solitons in Kerr media: CD helps to generate soliton families that exist under a variety of material and excitation conditions, and are *stable* against critical collapse.

VVK acknowledges support of FCT (Portugal) under grant PEst-OE/FIS/UI0618/2014.

### References
1. B. A. Malomed, D. Mihalache, F. Wise, and L. Torner, J. Opt. B **7**, R53 (2005).
2. D. Mihalache, J. Optoelectron. Adv. Mater. **12**, 12 (2010).
3. Y. V. Kartashov, B. A. Malomed, and L. Torner, Rev. Mod. Phys. **83**, 247 (2011).
4. L. Bergé, Phys. Rep. 303, 259 (1998).
5. E. A. Kuznetsov and F. Dias, Phys. Rep. 507, 43 (2011).
6. R. Y. Chiao, E. Garmire, and C. H. Townes, Phys. Rev. Lett. **13**, 479 (1964).
7. G. D. Montesinos, V. M. Perez-Garcia, and H. Michinel, Phys. Rev. Lett. **92**, 133901 (2004).
8. F. Lederer, G. I. Stegeman, D. N. Christodoulides, G. Assanto, M. Segev, and Y. Silberberg, Phys. Rep. **463**, 1 (2008).
9. Y. V. Kartashov, V. A. Vysloukh and L. Torner, Prog. Opt. **52**, 63 (2009).
10. Y. Sivan, G. Fibich, and M. I. Weinstein, Phys. Rev. Lett. **97**, 193902 (2006).
11. Y. V. Kartashov, B. A. Malomed, V. A. Vysloukh, and L. Torner, Opt. Lett. 34, 770 (2009).
12. M. L. Quiroga-Teixeiro and H. Michinel, J. Opt. Soc. Am. B 14, 2004 (1997).
13. S. K. Turitsyn, Theor. Math. Phys. **64**, 226 (1985).
14. O. Bang, W. Krolikowski, J. Wyller, and J. J. Rasmussen, Phys. Rev. E **66**, 046619 (2002).
15. B. A. Malomed, Soliton management in periodic systems, Springer-Verlag, Berlin, Germany (2005).
16. H. Sakaguchi, B. Li, and B. A. Malomed, Phys. Rev. E 89, 032920 (2014).
17. V. Galitski and I. B. Spielman, Nature 494, 49 (2013).
18. Y.-J. Lin, K. Jiménez-García, and I. B. Spielman, Nature 471, 83 (2011).
19. V. Achilleos, D. J. Frantzeskakis, P. G. Kevrekidis, and D. E. Pelinovsky, Phys. Rev. Lett. 110, 264101 (2013).
20. Y. Xu, Y. Zhang, and B. Wu, Phys. Rev. A 87, 013614 (2013).
21. Y. V. Kartashov, V. V. Konotop, and F. K. Abdullaev, Phys. Rev. Lett. 111, 060402 (2013).
22. V. E. Lobanov, Y. V. Kartashov, and V. V. Konotop, Phys. Rev. Lett. 112, 180403 (2014).
23. L. Salasnich, W. B. Cardoso, and B. A, Malomed, Phys. Rev. A 90, 033629 (2014).
24. K. S. Chiang, Opt. Lett. 20, 997 (1995).
25. K. S. Chiang, J. Opt. Soc. Am. B 14, 1437 (1997).
26. K. J. Blow, N. J. Doran, and D. Wood, Opt. Lett. 12, 202 (1987).
27. C. R. Menyuk, IEEE J. Quantum. Electron. QE-23, 174 (1987).
28. C. De Angelis, F. Matera, and S. Wabnitz, Opt. Lett. 17, 850 (1992).
29. J. M. Soto-Crespo, N. Akhmediev, and A. Ankiewicz, Phys. Rev. E 51, 3547 (1995).
30. N. Akhmediev, A. Ankiewicz, and R. Grimshaw, Phys. Rev. E 59, 6088 (1999).